\documentclass[slac_one]{revtex4}
\usepackage{graphicx}
\usepackage{fancyhdr}
\begin{document}

\title{Commissioning of the ATLAS Pixel Detector} %% Paper title goes here


\author{J.-F. Arguin on behalf of the ATLAS Pixel Collaboration}
\affiliation{Lawrence Berkeley Laboratory, 1 Cyclotron Road, Berkeley, CA 94720, USA}

\begin{abstract}
The ATLAS pixel detector is a high precision silicon tracking device
located closest to the LHC interaction point. It belongs to the first
generation of its kind in a hadron collider experiment. It will
provide crucial pattern recognition information and will largely
determine the ability of ATLAS to precisely track particle
trajectories and find secondary vertices. It was the last detector to
be installed in ATLAS in June 2007, has been fully connected and
tested in-situ during spring and summer 2008. It is currently in a
commissioning phase using cosmic-ray events. We present the highlights of the past and future
commissioning activities of the ATLAS pixel system.
\end{abstract}

\maketitle

\thispagestyle{fancy}


\section{Introduction}
\label{sec:intro}

The {\it A Toroidal LHC ApparatuS} (ATLAS) \cite{Atl} is a multi-purpose experiment that aims to explore the energy frontier of particle physics. It is made of successive detector layers, starting from the interaction points, of inner tracking detectors immersed in a 2T solenoidal magnetic field, electromagnetic and hadronic calorimetry, and muon detectors. The inner detector is composed of a silicon pixel detector, a silicon strip detector (the Semi-Conductor Tracker (SCT)) and the Transition Radiation Tracker (TRT). 

The design requirements for the innermost layer of the inner detector are largely driven by the hostile LHC environment. An irradiation dose of $10^{15}$ neq/cm$^2$ is expected by the end of the lifetime of the pixel detector. The pixel technology is well-suited for that purpose since it is intrinsically radiation hard. Furthermore the proton-proton collisions result in a very large amount of charged particles, approximately 1,000 tracks per event at the LHC design luminosity. The pixel detector achieves an occupancy per pixel of 10$^{-4}$ in this environment and is thus crucial for pattern recognition. Furthermore, since the pixel detector is located closest to the beamline and possesses a very good impact parameter resolution (12 $\mu$m in r-$\phi$ and 70 $\mu$m in the $z$ directions), it determines the capability of ATLAS to detect secondary vertices resulting from long-lived particles like B-hadrons contained inside $b$-jets.

\section{Pixel Detector Overview}
\label{sec:overview}

\begin{figure}
\begin{center}
\includegraphics[width=9 cm]{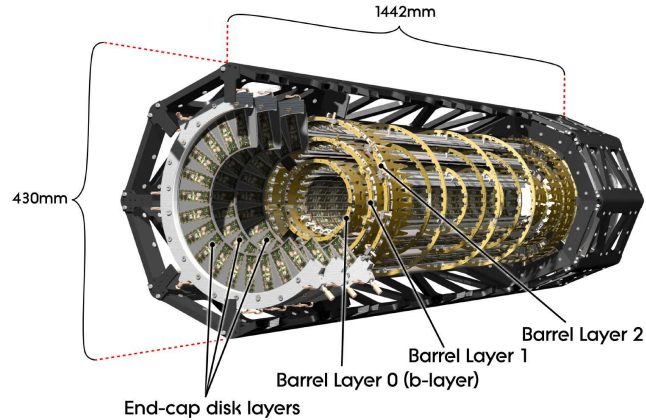}
\end{center}
\caption{Schematic view of the ATLAS pixel detector.}
\label{fig:pix}
\end{figure}

The general layout of the pixel detector is illustrated in Fig.~\ref{fig:pix}. The detector is made of three cylindrical layers in the barrel region (covering $|\eta|<1.9$) and three disks on each side of the end-cap region (covering $1.9 < |\eta| < 2.5$). The three barrel layers, the B-layer, L1 and L2, are located at radii of 5, 9 and 12 cm from the interaction point, respectively. 
%The detector volume is 1.4m in length and 0.2m in radius and the required total surface of silicon is 1.8m$^2$. The total amount of material is about 10\% $X_0$ in the transverse direction ($\eta=0$). 
%
The pixel detector is made of 1744 identical modules, each composed of a silicon sensor, 16 front-end chips, a controller chip and a flex printed circuit board. The silicon sensors are made of $n^+$ pixels on $n$ substrate and have a thickness of approximately 280 $\mu$m. A bonding technique is used to connect the sensors to the 16 front-end chips of the module with a pixel size of 50 $\times$ 400 $\mu$m in the $R-\phi$ $\times$ $z$ directions. Each pixel is connected to a charge sensitive feedback preamplifier and a discriminator with an adjustable threshold usually operated at $\approx$4 $ke$. The charge is collected at this threshold in less than 25 ns to allow unique bunch crossing identification of the hit. The signal is digitized  by calculating the time over threshold, which is proportional to the amplitude of the signal, using a 40 MHz timestamp (i.e. corresponding to the ATLAS clock speed). A typical signal remains over threshold for about 30 bunch crossings. Each pixel circuitry is equipped with a charge injection mechanism to tune and measure those parameters. The pixel hit, defined as the pixel address and the timing of the leading and trailing edges, is then transferred to the edge region of the front-end chip where it is stored in a hit buffer. %The length of this buffer as been shown to be sufficient to operate at full luminosity with the ATLAS level 1 trigger latency of 3.2 $\mu$s.
 The coincidence of the leading edge timing and the level 1 trigger is performed in this buffer.

 The accepted hits corresponding to the same bunch crossing are then transferred to the controller chip that collects the data from the 16 front-end chips and performs some initial event building. %and sends the data off-detector for further processing. 
The electrical signal is sent through micro-cables to optoboards \cite{optoboards} that perform the electrical to optical signal conversion. The optoboards are located on service panels \cite{Pix2} that bring the pixel services out of the inner detector volume. The data is then transferred off-detector to Read-Out Drivers (ROD) that performs further event building and data formatting and have a calibration capability. The ROD's are connected to the ATLAS data acquisition system through a 1.6 Gbit/s optical link. 

The total amount of power dissipated by the module electronics is approximately 10 kW. An evaporative cooling system (with coolant $C_3F_8$) is employed to operate the detector at a temperature $T\approx 0^\circ$C. %he modules are cooled through their support structure that is thermally conductive.

\section{System Test}
\label{sec:sys_test}
A system test of one pixel end-cap has been performed as a realistic test of the detector operations during the Fall of 2006. The program consisted in commissioning the setup, including cooling and readout, measurement of the analog performance of pixel modules and data-taking with cosmic rays. This program was completed successfully \cite{Pix}. A pixel efficiency of 99\% has been measured with an electronic noise level of $\approx 10^{-10}$.

\section{Pixel Package Integration and Installation}
\label{sec:integration}
During the Spring of 2007, the pixel detector was ready to be integrated with the service panels, the beampipe and the support structure to form the pixel package \cite{Pix2}. A connectivity test was performed in parallel to the mechanical work to verify that every part of the pixel package was functional before installation. This constituted also a commissioning activity since the complete detector was operated using the full readout chain for the first time. The pixel package was then inserted into the ATLAS inner detector volume in June of 2007. Electrical tests showed that the pixel package suffered no permanent damage as a result of the installation. At that stage, 99.7\% of the detector was known to be operational (0.1\% for module/optoboard failures, 0.2\% for pixel/front-end chip failures).

\section{Final Connection and Sign-Off}
\label{sec:connection}
After the SCT was connected and signed-off in Spring 2008, the connection of the pixel detector to its permanent services could begin. The electrical cables, which carry the high-voltage, low-voltage and module temperature signal, were first connected. The connection was verified in parallel with standalone programs using the pixel slow control system. Six modules with un-repairable high-voltage broken connections were found. The optical fibers were then installed. A test procedure using the full pixel data-acquisition chain monitored the light power output in both directions of the optolink communication. No on-detector failures were observed but several dead channels were detected in the off-detector VCSEL arrays. These channels keep dying to this day at the level of 1\% a week. Production of new VCSEL arrays is underway with improved safety.

The pixel package cooling pipes were then connected to the cooling plant (shared with the SCT). Leak tests showed that 3/88 cooling loops had leaks inside the pixel package and are thus inaccessible. Studies are underway to determine whether there are any long-term problems with the operation of these cooling loops. After the connections were completed, the plan was to complete the pixel sign-off by operating the cooling loops one at a time with the corresponding part of the detector. The sign-off was interrupted by a cooling plant accident involving the compressors that liquefy the fluid C$_3$F$_8$ when it returns to the cooling plant. Studies have shown that the cooling pipes inside the pixel package were not contaminated by this accident. The cooling plant has been fully cleaned and refurbished in a period of about 2$\frac{1}{2}$ months.

\begin{figure}
\begin{center}
\includegraphics[width=4cm]{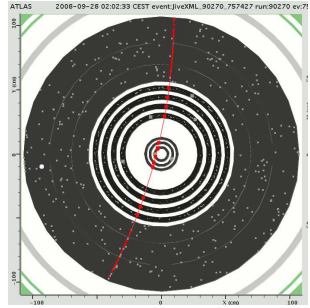}
\end{center}
\caption{A cosmic-ray track through the pixel detector (September 2008).}
\label{fig:pix2}
\end{figure}

The repairs of the cooling plant were completed in time to cool the detector during the bake-out of the ATLAS beampipe in early August 2008. The initial phase of commissioning started immediately after and is still underway at the time of writing this document. Calibration scans have been run on the whole detector to characterize and tune the optolinks and pixel analog electronics. Cosmic-ray events have been recorded with the pixel detector using trigger signal coming from the muon detectors and calorimeters. An example event is shown in Fig.~\ref{fig:pix2} when the solenoid field was on. Approximately 50,000 cosmic-ray events with pixel hits have been recorded to date.
\section{Future Plans}
The initial phase of commissioning of the pixel detector (without LHC beam) will continue until the Winter shutdown. The accumulated data will be used to tune and characterize the pixel detector and perform an initial alignment of the inner detector. We look forward to commission the pixel detector with LHC collisions in 2009.

\end{document}